\RequirePackage{ifthen}     
\RequirePackage{etoolbox}   
\RequirePackage{refcount}   
\RequirePackage{lastpage}   

\PassOptionsToPackage{hidelinks,breaklinks=true}{hyperref}

\documentclass[article,pdftex]{Definitions/mdpi}

\usepackage[T1]{fontenc}
\usepackage[utf8]{inputenc}
\usepackage{lmodern}

\usepackage{url}
\usepackage{textcomp}   

\makeatletter
\renewcommand{\leftcolDates}{}%
\renewcommand{\leftcolUpdateCitation}{}%
\renewcommand{\leftcolumnCorrstatement}{}%
\renewcommand{\leftcolumnCright}{}%
\renewcommand{\leftcolExternalEditor}{}%
\renewcommand{\leftcolumnUpdatelogo}{}%
\renewcommand{\cright}{}%
\makeatother

\providecommand{\pdfstringdefDisableCommands}[1]{}
\makeatletter
\fancypagestyle{fancy}{%
  \fancyhf{}%
  \fancyfoot[C]{\thepage}%
}
\makeatother

\makeatletter
\@ifundefined{linenomath}{%
  \newenvironment{linenomath}{}{}%
  \newenvironment{linenomath*}{}{}%
}{}%
\makeatother

\firstpage{1}

\pubvolume{1}
\issuenum{1}
\articlenumber{0}
\pubyear{2025}
\copyrightyear{2025}
\datereceived{ } 
\daterevised{ }
\dateaccepted{ } 
\datepublished{ } 
\hreflink{https://doi.org/} 

\Title{Accelerometer Measurements for Orbit and Gravity Recovery: Challenges and Benefits for the BepiColombo Mission}

\Author{Alireza HosseiniArani $^{1,2,\dagger}$\orcidA, Stefano Bertone $^{3,4}$, Daniel Arnold $^{2}$, William Desprats $^{2}$, Adrian J\"{a}ggi$^{2}$, and Nicolas Thomas$^{1}$}

\AuthorNames{Alireza HosseiniArani, Stefano Bertone, Daniel Arnold, William Desprats, Adrian J\"{a}ggi, and Nicolas Thomas}

\isAPAStyle{%
  \AuthorCitation{Lastname, F., Lastname, F., \& Lastname, F.}
}{%
  \isChicagoStyle{%
    \AuthorCitation{Lastname, Firstname, Firstname Lastname, and Firstname Lastname.}
  }{%
    \AuthorCitation{HosseiniArani, A.; Bertone, S.; Arnold, D.; Desprats, W.; J\"aggi, A.; Thomas, N.}
  }%
}

\address{%
$^{1}$ \quad Physics Institute, University of Bern, Switzerland; \\
$^{2}$ \quad Astronomical Institute, University of Bern, Switzerland; \\
$^{3}$ \quad University of Maryland College Park, CRESST II, USA; \\
$^{4}$ \quad Italian National Institute for Astrophysics (INAF), Astrophysical Observatory of Torino, Italy; \\
}

\firstnote{Institute of Geodesy (IfE), Leibniz University Hannover, Germany}

\abstract{%
The European Space Agency's BepiColombo mission continues its pioneering voyage to Mercury, the innermost planet of the Solar System. Among the advanced instruments onboard the Mercury Planetary Orbiter (MPO), one of the two spacecraft that comprise the BepiColombo mission, is the Italian Spring Accelerometer. The instrument's primary scientific goals are closely linked to the Mercury Orbiter Radio-Science Experiment. Together, these instruments aim to provide valuable data on the spacecraft's orbit, as well as Mercury's gravity field and internal structure.    
This simulation study examines how modeling and parametrization of accelerometer measurements affect orbit and gravity field recovery, and it explores strategies to overcome the challenges associated with the co-estimation of all parameters. In it, we evaluate the accuracy of the integrated retrieval of spacecraft orbit, gravity field, and accelerometer parameters under different noise levels and varying observation geometries during the mission.
To achieve these goals, we propagate the orbit of MPO and simulate Doppler observations and accelerometer measurements based on the available noise models. We consider two scenarios by including either an optimistic or a more realistic accelerometer noise level.  
Our results indicate that postponing the estimation of accelerometer biases until a preliminary gravity field is established helps prevent gravity field mismodelings from being absorbed into the accelerometer solution. The daily estimation of bias was found to be essential. Using one year of Doppler tracking data and applying Kaula regularization, we demonstrate the potential to recover Mercury's gravity field up to degree and order $40$. Improved recovery of low-degree coefficients was observed under optimistic noise assumptions. Errors in cross-track bias estimation increased sharply when the $\beta$-Earth angle dropped below $45^\circ$, corresponding to the degraded Doppler observability periods. Orbit determination achieved centimetre-level accuracy in the radial direction and metre-level accuracy in the along- and cross-track directions, with a substantial rise in cross-track orbit errors observed during periods of reduced Doppler observability.
}


\begin{document}
\pagestyle{fancy}
\thispagestyle{fancy}

\makeatletter
\ifdefined\nolinenumbers
  \nolinenumbers
\fi
\makeatother

\section{Introduction}\label{sec:intro}

The BepiColombo mission~\citep{BENKHOFF2010} is a collaborative effort between the European Space Agency (ESA) and the Japan Aerospace Exploration Agency (JAXA) designed to explore Mercury, the innermost planet of the Solar System. Launched in October 2018, BepiColombo aims to provide unprecedented insights into Mercury's gravity field, composition, geology, atmosphere, magnetosphere, and its interaction with the solar wind. The mission consists of two spacecraft: the Mercury Planetary Orbiter (MPO), developed by ESA, and the Mercury Magnetospheric Orbiter (MMO/Mio), developed by JAXA. Together, these spacecraft will conduct a comprehensive study of Mercury's surface and environment, helping to answer fundamental questions about the planet's formation and evolution, as well as offering clues about the broader processes that shaped the Solar System. 

Onboard MPO, the Italian Spring Accelerometer~\cite[ISA,][]{ISA2001} and the Mercury Orbiter Radio-science Experiment~\cite[MORE,][]{MORE2001} are the two main instruments designed to address BepiColombo's scientific goals in geodesy, geophysics and fundamental physics. These two instruments, together with BepiColombo Laser Altimeter~\citep[BELA,][]{Thomas2007} and the integrated Suite for the Imaging and Spectroscopic Investigation System~\citep[SIMBIO-SYS,][]{Cremonese2020}, 
contribute to the determination of the spacecraft orbit and of the shape and gravity field of Mercury~\citep{Marabucci2013, HOSSEINIARANI2021}. They will also provide crucial experimental constraints to models of the planet's internal structure~\citep{Spohn2001} and test theories of gravity~\citep{PPN}. 


Among measurements delivered by MORE, the $2$-way range-rate is measured from the Doppler shift of an electromagnetic wave transmitted from the Earth, received by the spacecraft and coherently retransmitted back to Earth. 
The key instrument is the Ka-band transponder onboard MPO. The Ka-band link is coherently converted from $34$~GHz (uplink) to $32$~GHz (downlink). 
The uplink signal is also modulated by a wideband pseudo-noise ($50~MHz$ bandwidth), which is fully demodulated onboard and then retransmitted to the ground by phase modulating the downlink carrier. This measurement is expected to have an accuracy of $1.5~\mu$m/s at $1000$~s integration time~\citep{Benkhoff2021}. 
In addition to the Ka-band system, MORE also employs an X-band link, which enables plasma noise mitigation—particularly important when the spacecraft is near the Sun—by allowing differential frequency measurements that isolate and reduce the dispersive effects of the solar plasma~\citep{Iess2018}.
\citet{DiStefano2022AGU} detail important in-flight calibration efforts for the MORE experiment undertaken during the cruise phase, emphasizing the importance of calibrating dispersive plasma noise components and utilizing water vapor radiometers to correct tropospheric delays.

The proximity of Mercury to the Sun makes the accurate knowledge of the non-gravitational forces acting on MPO critical for the quality of orbit determination. To overcome modeling limitations, a three-axis high-sensitivity accelerometer is placed on board MPO to give accurate information on the non-gravitational accelerations. ISA detects the displacements of a proof mass due to perturbing accelerations by means of capacitive transducers in a bridge configuration, followed by a low-noise amplifier. It then uses them to estimate the non-gravitational accelerations acting on the MPO structure. 
ISA's key role is to calibrate non–gravitational perturbations acting on the probe, \textit{i.e.}, non–conservative accelerations related to direct radiation emitted by the Sun and re-radiated by Mercury, from the list of quantities to be estimated.


Thanks to BepiColombo’s near-polar orbit and accurate tracking, the gravity field will be mapped globally, in contrast to MESSENGER’s northern-hemisphere-biased mapping. Notably, gravity anomalies and mass concentrations in Mercury’s southern hemisphere – which remained poorly resolved by MESSENGER – will be accurately determined by BepiColombo for the first time~\citep{Benkhoff2021}. 
By the end of its nominal one-year orbital mission, MPO’s tracking data (range, Doppler) in combination with ISA will yield a refined gravity model of Mercury. Formal covariance analyses predict substantially smaller uncertainties in Mercury’s low-degree harmonics and tidal response compared to MESSENGER-era knowledge~\citep{Iess2021}. Similar studies on the orbit determination conclude that accuracies of $0.1-1$ m in the radial position recovery seem attainable~\cite{Genova2011}. 
In addition to geodetic studies, the accurate positioning of the MPO in the hermean frame will also be critical for the appropriate referencing of the laser altimetric measurements and the images from the high-resolution camera.

Several simulation studies have investigated the expected performance of BepiColombo's Mercury Orbiter Radio Science Experiment (MORE), focusing on the recoverable gravity field resolution and orbit determination accuracy. 
\citet{Milani2001} first demonstrated that one year of BepiColombo tracking data combined with onboard accelerometry could map Mercury's gravity field to at least spherical harmonic degree 25.
Orbital errors in the radial direction were found to be well below 1~m, confirming that decimeter-level accuracy was achievable. 
Importantly, their study highlighted the necessity of including arc-wise accelerometer bias estimation to absorb non-gravitational perturbations. 
\citet{Milani2001} also discuss the dependency of accelerometer bias on temperature, examining its temporal variation. The authors argue that, due to the strong thermal insulation of the spacecraft, short-term variations in heat flow lead to only minor changes in internal temperature. 
\citet{Iess2009} refined the mission's radio science design, presenting expected Doppler and range measurement accuracies and showing that tracking data combined with the calibrated accelerometer data could support gravity field recovery up to degree 25. 
Their analysis confirmed that accelerometer biases must be carefully estimated to avoid contaminating the gravity solution.
\citet{ABC2016} provided a covariance analysis, demonstrating reliable gravity recovery up to degree $25$ with a signal-to-noise ratio of $10$ and confirming centimeter-level accuracy for the radial component of the MPO orbit. They show that the along-track orbital errors were larger but remained within the meter range. 
They confirmed that accelerometer biases in all three axes could be estimated per arc to within a few $\times10^{-9}$~m/s$^2$ under realistic mission conditions. They also concluded that the non-calibrated component at spacecraft orbital period, together with the residual effects after calibration at lower frequencies, causes the true errors to be, in general, worse than formal ones.
Another simulation study by \citet{Iess2021} confirmed that the BepiColombo mission is expected to recover Mercury's static gravity field up to degree 35 without regularization, and up to 45 with Kaula constraints. 
The same study reported orbit determination errors of approximately 10~cm in the radial direction, and on the order of 1–10~m and $\sim$1~m in the along-track and cross-track directions, respectively.

%
Accelerometer biases and drifts were solved in each orbit arc using a multi-step estimation strategy, with residual accelerations effectively suppressed below the sensitivity of the tracking system.

These studies show that achieving BepiColombo’s ambitious orbit and gravity field recovery objectives requires overcoming several challenges in data processing and modeling. A primary concern is the proper calibration of ISA. While the accelerometer significantly mitigates non-gravitational perturbations such as solar radiation pressure and thermal thrust, any unmodeled biases or scale factor errors can lead to systematic errors in orbit determination and, as a consequence, gravity field estimation \cite{DeFilippis2024}.
In particular, accelerometer biases correlate with gravitational accelerations, thereby corrupting the estimation of Mercury’s gravitational field coefficients or the spacecraft orbit \cite{DelVecchio2023,DeFilippis2024} if not properly accounted for. Simulation studies have shown that accelerometer biases must be treated as solve-for parameters in the orbit determination filter, and in some cases, may require empirical modeling or periodic in-flight calibration strategies. Such calibration could include exploiting known force-free periods or maneuver-related events where true accelerations are well characterized \cite{DelVecchio2023}. Operational factors further complicate this calibration. For instance, the Mercury Planetary Orbiter (MPO) will undergo frequent momentum wheel desaturation maneuvers, which introduce brief but significant thrust events. ISA can detect and quantify these transient accelerations, provided its bias and noise characteristics are well constrained \cite{Zannoni2021}. Moreover, the spacecraft is subject to low-level outgassing events, as observed during BepiColombo's Venus flybys, where ISA captured unexpected acceleration spikes \cite{DelVecchio2023}. Without the accelerometer, such events would not be properly characterized, introducing errors in the orbit solution.

The goal of this study is to further assess the impact of the ISA on orbit determination and gravity field recovery. We address the challenges introduced by accelerometer biases on gravity recovery and explore potential strategies to overcome this. We then investigate the accuracy of the recovery of the orbit, gravity field and accelerometer parameters, assuming different noise levels and under different observation geometries.
We conduct a closed-loop simulation, propagating the orbit to generate synthetic accelerometer measurements and Doppler data. Orbit estimations are performed in daily arcs, solving for the initial state vector of each arc, gravity field coefficients, and accelerometer parameters. 
We use the Bernese GNSS Software~\citep{Dach2015} for simulation and orbit determination. While this software is widely applied to precise orbit determination of Earth-orbiting satellites using onboard GNSS data, its planetary extension has also been used in lunar geodesy~\citep{bertone2021} and for simulating orbit determination in planetary missions~\citep{DESPRATS2023,desprats2024thesis,Desprats2025}.

The modeling of the spacecraft orbit, radio science instrument, and accelerometer is presented in Section~\ref{sec:modeling}. Section~\ref{sec:strategy} outlines the orbit determination strategy, including key challenges and potential solutions. The results are presented in Section~\ref{sec:resultsSec}, followed by a discussion in Section~\ref{sec:diss}, where the findings are interpreted in the context of previous studies.

\begin{figure}[H]
    \centering
        \includegraphics[width=0.95\columnwidth]{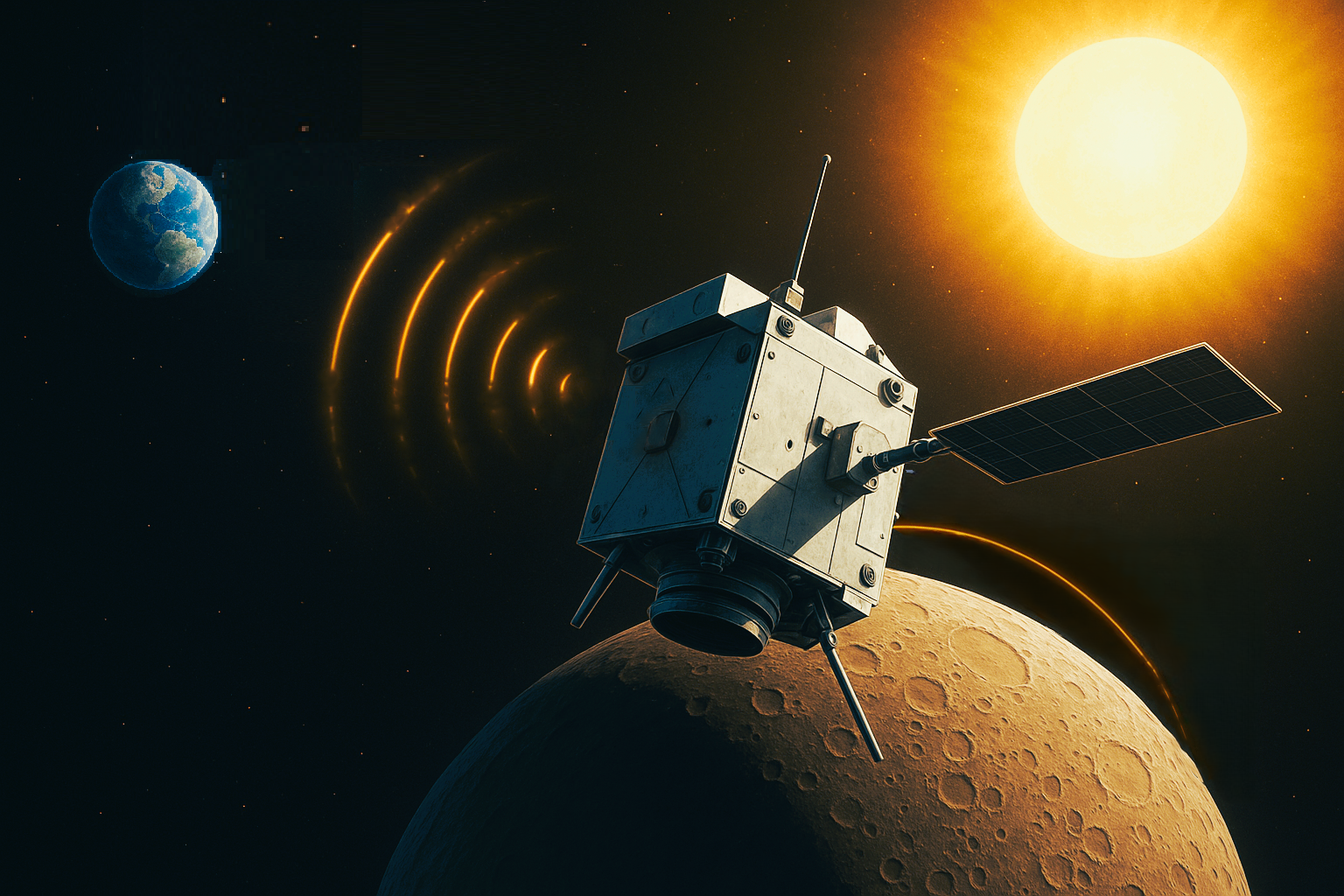}
    \caption{Artistic visualization of the spacecraft around Mercury; orbit determination is done using the two-way radio tracking signal. Image generated using OpenAI's ChatGPT (DALL·E), June 2025.}
    \label{fig:tracking}
\end{figure}

\section{Simulation Setup}\label{sec:modeling}

\subsection{Orbit Model}\label{sec:modelingOrb}
We use a full-force model including the gravitational and non-gravitational forces to
compute the orbit of the spacecraft over a one-year period by propagating daily arcs from the ESA-provided nominal orbits~\citep{Jehn2015}. 
To ensure consistency in the comparison, the same gravity field and similar dynamical assumptions as those described in~\citep{Jehn2015} were used. 
In particular, we include Mercury's gravity field HGM005~\citep{Mazarico2014b} up to degree and order (d/o) 50, third-body gravity perturbations by the Sun and planets as point masses, and tidal forces exerted by the Sun on Mercury. We also apply common
relativistic corrections affecting the spacecraft orbit, including the Schwarzschild, Lense-Thirring, and de Sitter corrections. We model the Solar Radiation Pressure and Planetary Radiation Pressure due to reflected and emitted infrared radiation (IR) by Mercury. To consider the effect of non-gravitational force
on the spacecraft consistently with~\citep{Jehn2015}, we use a 33-plate macromodel of MPO, including both visible and IR optical properties~\citep{NickPrivate}. 
The resulting orbit was validated against the reference solution~\citep{Jehn2015}, with the differences found to be $\sim 15$ cm at most in each daily arc~\citep{hosseinearani2020independent}.

\subsection{Radio Science Model} \label{sec:modelingDop}
An artistic visualisation of the BepiColombo mission is shown in Figure~\ref{fig:tracking}. 
We simulate two-way K-band Doppler tracking measurements considering the visibility conditions of spacecraft with respect to Earth. 
The visibility conditions depend on the $\beta$-Earth angle, defined as the angle between the spacecraft's orbital plane and the line-of-sight vector from Earth to the spacecraft. This angle influences how the spacecraft’s orbit appears from Earth and plays an important role in the sensitivity of radio-tracking data to the state vector of deep-space missions.

Following~\cite{IESS2001} and assuming nominal accuracy performances of the transponder in the Ka-band, we model tracking measurements errors with a Gaussian white noise with a standard deviation of $1.5$~mHz $@10$~s, hich is added to the simulated range-rate~\citep[similar to][]{ABC2016}. 

\subsection{Accelerometer Model} \label{sec:accModel}
Within the orbit reconstruction process, the explicit modeling of the non-conservative forces (\textit{e.g.}, solar radiation pressure) is replaced by simulated accelerometer measurements. To simulate the accelerometer measurements at each observation time, we first calculate the total non-gravitational acceleration acting on the spacecraft, then we add the expected accelerometer noise.
The ISA error model~\cite{Iafolla2010,ABC2016} can be summarized by the following main terms: the spacecraft's orbital period or resonant term (approximately $2.3$ hours), other systematic components, Mercury's orbital period or main thermal term (approximately $88$ days), and a random component.

The temperature variations (peak-to-peak) estimated for the mentioned thermal effects are about $25^\circ C$ in the case of the long-period effects and about $4^\circ C$ in the case of the effects at MPO orbital period \citep{Iafolla2007}. 
Therefore, in order to avoid these effects, a strong thermal insulation of the spacecraft and an active thermal control~\citep{iafolla2006} will be used to attenuate their impact on the accelerometer readings. The results of the lab experiments have presently shown an attenuation factor of about $700$, both at the Mercury revolution period and at the MPO orbital period~\citep{Lucchesi2006}, which results in thermal effects to only cause minor changes in internal temperature.


The measurement bandwidth of ISA is between $3 \times 10^{-5} Hz$ and $3 \times 10^{-1} Hz$. 
ISA performs better at frequencies above $1 \times 10^{-3} Hz$, while at lower frequencies, the measurement suffers from a larger noise resulting in a bias in the accelerometer readings~\citep{Lucchesi2006}. 
In general, the accelerometer measurements along the $i^{th}$ axis of the accelerometer frame can be written as 

\begin{linenomath}
\begin{equation}
A_{i, meas} = B_i + S_i \times A_{i, true} + N_i \; ,
\label{eq:ACC}
\end{equation}
\end{linenomath}

where $A_{i, true}$ is the true non-gravitational acceleration acting on the spacecraft along the $i^{th}$ axis, $B_i$ is the accelerometer bias, $S_i$ is the scaling factor, $A_{i, meas}$ is the accelerometer measurement along the same axis, and $N$ is the accelerometer random noise~\citep{ISA2001}. 
The scaling factor of an accelerometer refers to how its output changes in response to a known input acceleration. For most high-precision accelerometers, this scaling factor can be measured accurately in controlled calibration setups before launch and adjusted periodically during the mission by using known reference signals or maneuvers. Scaling factor calibration is usually stable over time and can be adjusted with high precision because it relies on known physical principles and predictable response to input changes.
\citet{Astrua_2023} details the calibration procedures for ISA and highlights that the scale factor calibration achieves a relative accuracy of 240 parts per million (ppm), emphasizing the stability and precision attainable through calibration based on known physical principles.
The bias of an accelerometer, on the other hand, is the output it reads when there is theoretically zero acceleration. Bias tends to be more variable and susceptible to environmental factors. Unlike scaling factors, biases are more difficult to calibrate in-flight and can drift unpredictably, making precise determination and long-term stability challenging~\citep{Touboul2014}.
For this reason, in this study, we ignore the scaling factor and focus on the determination of the accelerometer bias. 


\citet{Milani2001} discuss that the accelerometer bias consists of a temperature-dependent component and a constant offset, modeled as $B = dT + c$. While the coefficient $d$ is well characterized from ground measurements as approximately $5 \times 10^{-4}~\text{cm/s}^2/\text{K}$, they emphasize that the absolute internal temperature cannot be accurately determined in space, and the offset term $c$ is inherently different from that obtained in ground experiments. As a result, they propose a digital calibration approach in which $B$ is treated as an additional parameter in the least-squares estimation. They also show that the bias can reasonably be assumed constant over short orbital arcs, making such modeling feasible within typical orbit determination intervals.

\begin{figure}[H]
\isPreprints{\centering}{} 
    \includegraphics[width=0.95\columnwidth]{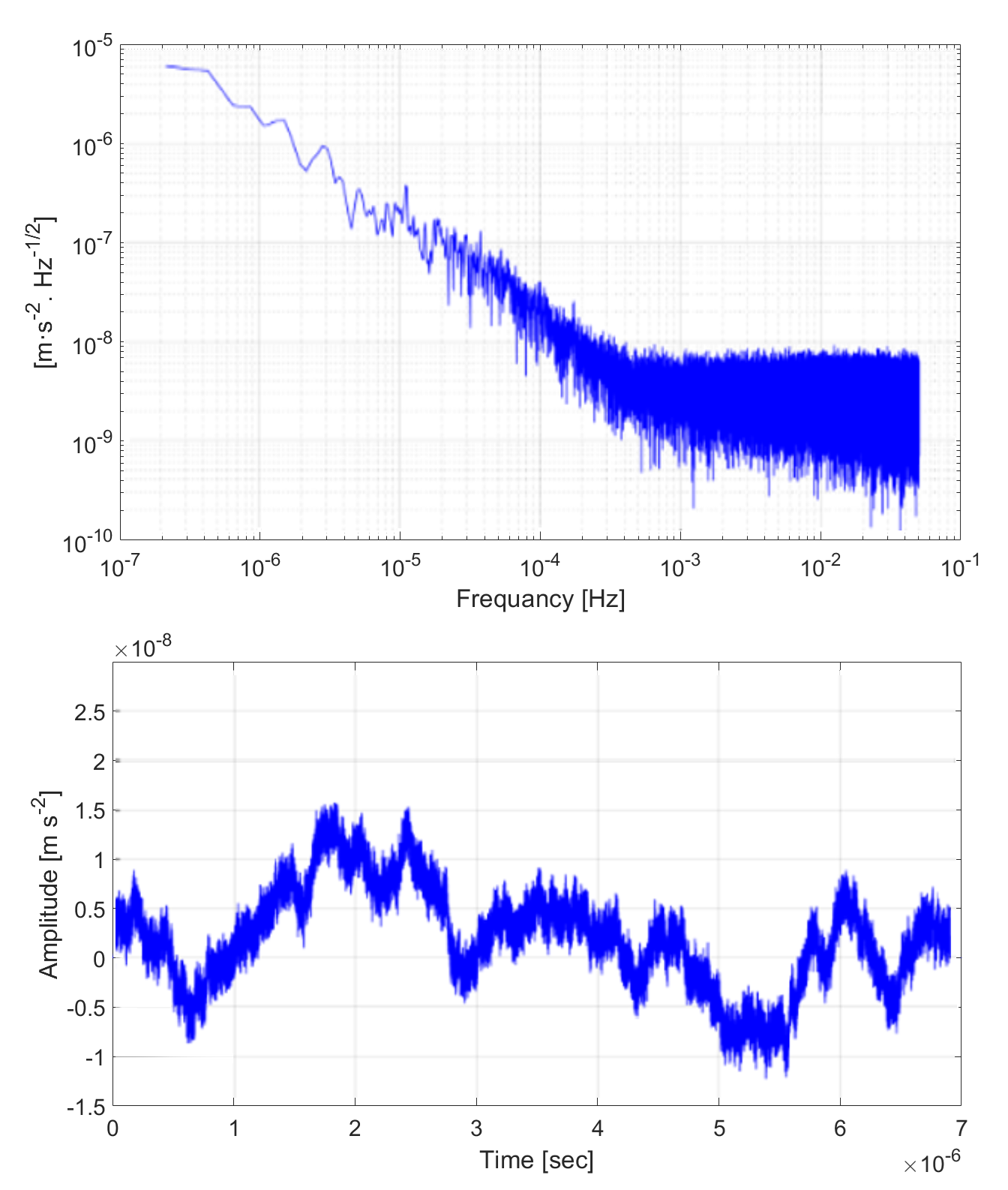}
    \caption{Simulation of the ISA accelerometer noise in frequency (top) and time (bottom) domains.}
    \label{fig:ISAnoiseGen}
\end{figure}

Figure~3 of \cite{ABC2016} and Figure~1 of \cite{alessi2012acc} present the expected noise of the ISA accelerometer in the frequency and time domains, respectively.  
We simulate a similar noise, which is shown in Figure~\ref{fig:ISAnoiseGen}. This simulated noise is equivalent to a random noise of $3 \times 10^{-9}~m/s^2$ and a bias amplitude in the order of $1 \times 10^{-8}$ to $2 \times 10^{-8}~m/s^2$
\citep{ISA_team_private_comm}. 
We consider the random noise of $3 \times 10^{-9}~m/s^2$ to be our optimistic scenario. Considering the challenges that ISA might face around Mercury, we also assume another case in which we have a similar level of bias and white noise of $1 \times 10^{-8} m/s^2$. We show our simulated noise for this ``realistic scenario'' in Figure~\ref{fig:ISAnoiseGen}. 

\subsection{Modeling of the Desaturation Maneuvers}

Onboard reaction wheels contribute to the pointing accuracy needed for instruments onboard MPO. The angular momentum accumulated by the reaction wheels has to be regularly reset, which results in periodic desaturation maneuvers. During such maneuvers, which are foreseen approximately every $12$ hours~\citep{Iess2021}, the unbalanced thrusters will produce a large along-track $\Delta V$. The knowledge of the $\Delta V$ associated with desaturation maneuvers is possible to a level of $2-5 \%$ or $1.2-3$ mm/s~\citep{Iafolla2010}. To consider the desaturation maneuvers, we estimate a velocity pulse with the same uncertainties/constraints every $12~h$ during the orbit reconstruction~\citep{Jaeggi2006}. 

\section{Strategy and Approach} \label{sec:strategy}

We integrate the orbit of MPO with a realistic (gravitational and non-gravitational) force model to produce synthetic Doppler observations and accelerometer measurements. We take the initial conditions of each daily arc from the simulation, add a random offset to it and use them as a priori values for the estimation. The standard deviation of the error on the initial positions and velocities in each direction is $10$~m and $0.01$~m/s, respectively, which is consistent with~\citet{Iess2021}. Then, we use a degraded force model (truncated gravity field + synthetic accelerometer measurements) with errors on the initial state vector as the “a priori” knowledge of the orbit. 

\unskip

We base our recovery of the MPO orbit on a least-squares adjustment~\citep{Beutler2010} of simulated Doppler observables and simulated accelerometer measurements. We estimate orbits in daily arcs. Each arc has its own initial conditions and is independent from the others. The solved-for parameters are divided in two groups: the local parameters that are defined and estimated for each arc, (\textit{e.g.}, initial state vectors and accelerometer biases), and global parameters, (\textit{i.e.}, the gravity field coefficients). 

We use HGM005~\citep{Mazarico2014b} up to d/o 50 as the ``true'' Mercury's gravity field in our simulation. As different available MESSENGER-based gravity field solutions start deviating from each other at d/o 10 (see, \textit{e.g.}, 
Fig. 2 of \cite{Imperi2018}), we assume in the recovery step that the true gravity field is only known up to d/o 8. Consequently, in the gravity recovery procedure, HGM005 model is employed as the a priori field but truncated at d/o 8. 
For simplification, we assume the availability of three ground stations~\citep{Montagnon2021} for spacecraft tracking and analyze the combined recovery of the gravity field, orbit, and accelerometer parameters. 

We perform several orbit reconstruction tests using daily arcs with noise-modulated Doppler data. 
The orbit is recovered by solving for the initial state vectors of the arcs, a process we refer to as the orbit fit, which represents the local-level iteration. On top of this, we aim to reconstruct the gravitational field used in the simulation, initially perturbed at the orbit fit stage. 
For this purpose, we combine normal equations from the daily arcs over one year of mission and solve for the coefficients of the gravity field and spacecraft orbit parameters simultaneously. 
This constitutes a global iterative process, where in each iteration we update the a priori gravity field with the solution obtained from the previous global step. Within each global iteration, the local orbit fits are recomputed using the updated field, and the new set of normal equations is accumulated for the subsequent global solve.

To absorb mismodelings, we estimate pseudo-stochastic pulses~\citep{Beutler2010} every $72$ min with a loose constraint of $1 \times 10^3~m/s$ at the early stages of the gravity field recovery. We tighten the constraint in the later stages to $1 \times 10^{-3}~m/s^2$. At the initial iteration of the orbit fit, a considerable number of arcs exhibit large Doppler residuals and large orbit discrepancies at overlaps. This is because orbit parameters cannot absorb the large force model deficiency which comes from non-estimated gravity field coefficients in this step, despite the loosely constrained pulses. If not excluded, these arcs compromise the orbit and gravity solution in subsequent iterations. To address this, we exclude arcs with orbit overlaps greater than $5$ km and Doppler residuals exceeding $5$ Hz. This procedure eliminates over $50\%$ of the arcs at the first iteration but it allows to improve the gravity solution from the significantly truncated a priori gravity field. Importantly, we notice that the removal of problematic arcs is only necessary during the first iteration of gravity recovery. However, subsequent orbit-fitting steps based on the newly estimated gravity field solution still need $6-10$ iterations to achieve convergence.
\begin{figure}[H]
\begin{adjustwidth}{-\extralength}{0cm}
\centering
    \includegraphics[width=17.0cm]{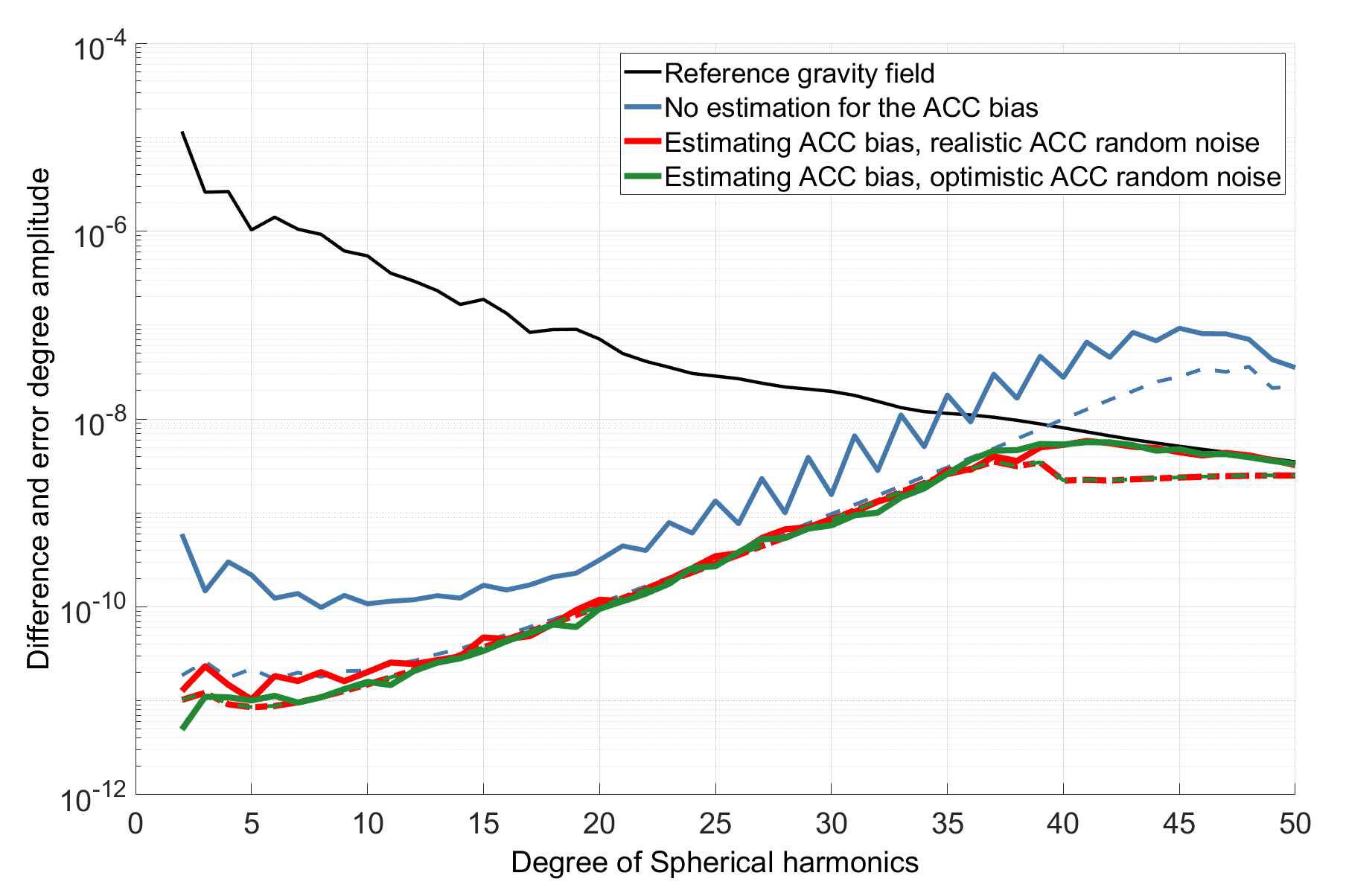}

        \caption[gravity field recovery]{Difference and error degree amplitude using one year of MPO's 2-way Doppler observations. The blue curve shows the recovered gravity field with biased accelerometer data; further recovery of the gravity field during the nominal mission, by co-estimating accelerometer biases when assuming a realistic (red) or optimistic (green) accelerometer noise model. Differences are mainly visible at low degrees, where the green curve matches formal uncertainties. A Kaula regularization constraint was applied from d/o 40 upwards, using a standard deviation of $\sigma_n = 10^{-4} / n^2$ for each degree $n$.}
        \label{fig:gravRec}
\end{adjustwidth}
\end{figure}


Our simulations indicate that estimating accelerometer biases when starting from a poor a priori gravity field (d/o 8 in our case) may result in a degraded recovery because of correlations between accelerometer and gravity field parameters. For instance, when doing so in the early iterations (and especially without a strong constraint to 0), accelerometer parameters are estimated to values orders of magnitude higher than the ``true'' ones, often resulting in diverging solutions. We concluded that the best approach would be to strongly constraining accelerometer parameters to zero until orbits and the gravity field are improved to the extent possible with biased accelerometer information. Therefore, during the initial iterations of gravity recovery, we exclude the accelerometer parameters from the set of solve-for parameters, assuming them to be zero, despite their non-zero simulated values. Once this preliminary gravity field solution has converged (after $\sim 4$ iterations, see the blue curve in Fig~\ref{fig:gravRec}), we proceed to co-estimate the accelerometer biases along with the orbital parameters and gravity field coefficients. 

In Section~\ref{sec:resultsSec} we show that, when starting from a sufficiently accurate gravity field, simultaneous recovery of gravity, orbital parameters, and accelerometer biases leads to further improvements of the overall solution.
Our simulations indicate that when starting from a perfect a priori gravity field (identical to the one used in the simulation), stacking the biases over longer arcs improves the accuracy of the bias recovery~\citep{hosseiniarani2022COSPAR, desprats2024thesis}. Here, stacking refers to solving for a single constant bias parameter shared across multiple arcs, rather than estimating separate biases for each arc individually. However, in more realistic scenarios where the starting point is a previously estimated (non error-free) gravity field, stacking the accelerometer biases over longer arcs does not necessarily enhance the recovered gravity field or the accelerometer biases, and may even degrade the solution.
Our findings suggest that, once the gravity field solution converges with no accelerometer biases, estimating daily accelerometer biases enables additional improvements by absorbing remaining mismodelings in both the accelerometer and OD model.
It is important to note that 
improving the gravity field without estimating accelerometer biases in the early iterations is a key factor for the successful convergence of the problem in the final iterations.

\section{Results} \label{sec:resultsSec}
\subsection{Recovery of the Gravity Field: Iterations to Convergence without Estimating Accelerometer Biases}

Based on the available model, several orbit reconstruction and gravity recovery tests have been conducted. Figure~\ref{fig:gravRec} presents a comparison of the results from a gravity recovery test using one year of MPO's 2-way Doppler observations under varying assumptions for accelerometer bias recovery.

As mentioned in Section~\ref{sec:strategy}, during the initial stages of gravity field recovery, the accelerometer parameters are excluded from the set of estimated parameters and are assumed to be zero, even though their simulated values are non-zero. 
The blue curve in Figure~\ref{fig:gravRec} shows the resulting gravity field solution after several iterations to convergence under this assumption.

\subsection{Recovery of the Gravity Field: Incorporating the Accelerometer Biases}

Once a preliminary gravity field solution is obtained after iterations without estimating for the biases, the accelerometer biases are introduced into the estimation process, along with the spacecraft’s orbital parameters and the gravity field coefficients.

The accelerometer biases vary randomly with an RMS of $1 \times 10^{-8}~ms^{-2}$ in our simulations, then estimated from a zero a priori alongside the coefficients of the gravity field, by incorporating either (1) 
the more realistic accelerometer noise of $1 \times 10^{-8}~\text{ms}^{-2}$, or (2) the more optimistic accelerometer noise level ($3\times 10^{-9}~\text{ms}^{-2}$), as defined in Section~\ref{sec:accModel}. In both scenarios, the accelerometer biases are treated as daily parameters
, achieving an accuracy better than the initial bias amplitude. A daily bias estimation is expected to provide a more realistic representation than long-arc biases, since accelerometer parameters are expected to vary on a daily basis (see Section~\ref{sec:accModel}), thus resulting in a better data fit. 

Table~\ref{biasRec} presents the RMS errors of the daily estimated accelerometer biases at the final iteration, shown for different directions and under different assumptions about accelerometer noise (see Section~\ref{sec:accModel}).
The results show that our final daily estimates are especially accurate and stable in the along-track direction (a few $\%$ errors), while also leading to reliable improvements in the radial direction. However, estimating cross-track accelerometer biases proved challenging.


\begin{figure}[htbp]
\begin{adjustwidth}{-\extralength}{0cm}
\centering
    \includegraphics[width=19.0cm]{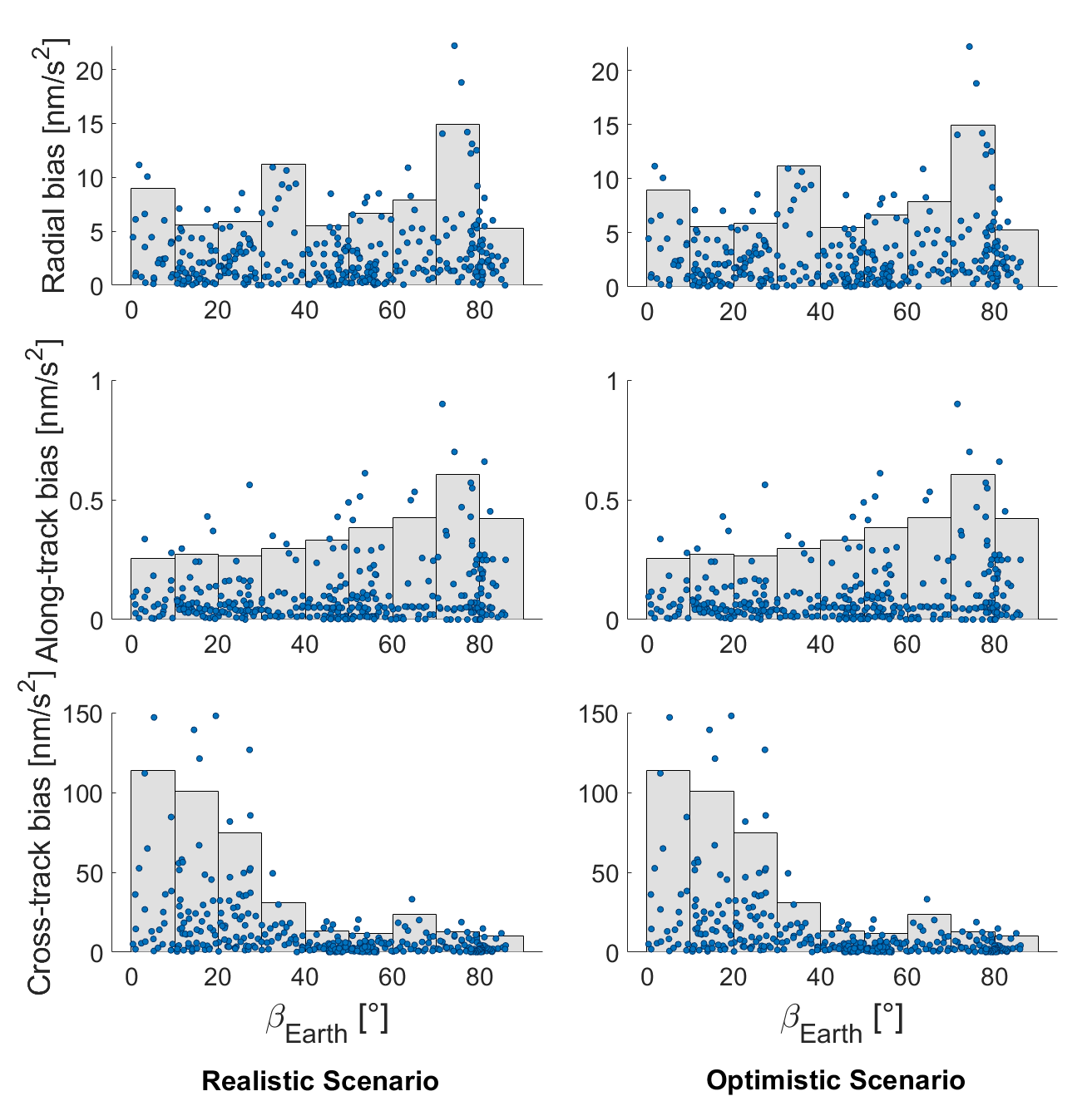}
    \caption[Accelerometer-bias RMS vs beta angle]{Daily RMS error of the radial (top), along-track (middle), and cross-track (bottom) accelerometer biases as a function of the $\beta$-Earth angle (see Section~\ref{sec:modelingDop}). $3\sigma$ bins are plotted on top of the scatter values to highlight the dependency on the $\beta$-Earth angle. The left column corresponds to the realistic scenario, while the right column corresponds to the optimistic scenario.}
    \label{fig:bias_beta}
       \end{adjustwidth}
\end{figure}

\begin{table}[htbp]
\caption[RMS error of acc bias]{RMS of errors in the recovered daily accelerometer bias parameters, where the simulated true biases vary randomly with an RMS of $1 \times 10^{-8}$\,m\,s$^{-2}$.
\label{biasRec}}
	\begin{adjustwidth}{-\extralength}{0cm}
		\begin{tabularx}{\fulllength}{lCCC}
			\toprule
			\textbf{Scenario}	& \textbf{Radial bias}	& \textbf{Along-track bias}     & \textbf{Cross-track bias}\\
			\midrule
            \multirow{2}{*}{Realistic scenario}          & \multirow{2}{*}{$6.7 \times 10^{-9}~m/s^2$ } & \multirow{2}{*}{ $3.7 \times 10^{-10}~m/s^2$ } & \multirow{2}{*}{ $4.4 \times 10^{-8}~m/s^2$ } \\
 &   &  &   \\
			\midrule
\multirow{2}{*}{Optimistic scenario}           & \multirow{2}{*}{$4.0 \times 10^{-9}~m/s^2$ } & \multirow{2}{*}{$1.7 \times 10^{-10}~m/s^2$ } & \multirow{2}{*}{ $2.4 \times 10^{-8}~m/s^2$ }\\
  &   &  &   \\
			\bottomrule
		\end{tabularx}
	\end{adjustwidth}
\end{table}

The red and green curves in Figure~\ref{fig:gravRec} show our gravity field solutions obtained after the (realistic and optimistic, respectively) accelerometer biases are incorporated into the set of estimated parameters.
Our results indicate an improvement in the accuracy of the recovered gravity field when the accelerometer bias is accounted for. As expected from long period orbital errors, the main difference between the red and green solutions is visible in the lower degrees. The gravity recovery results presented here are overall consistent with previous findings, as reviewed in Section~\ref{sec:intro}.


\begin{figure}[htbp]
\begin{adjustwidth}{-\extralength}{0cm}
\centering
    \includegraphics[width=18.0cm]{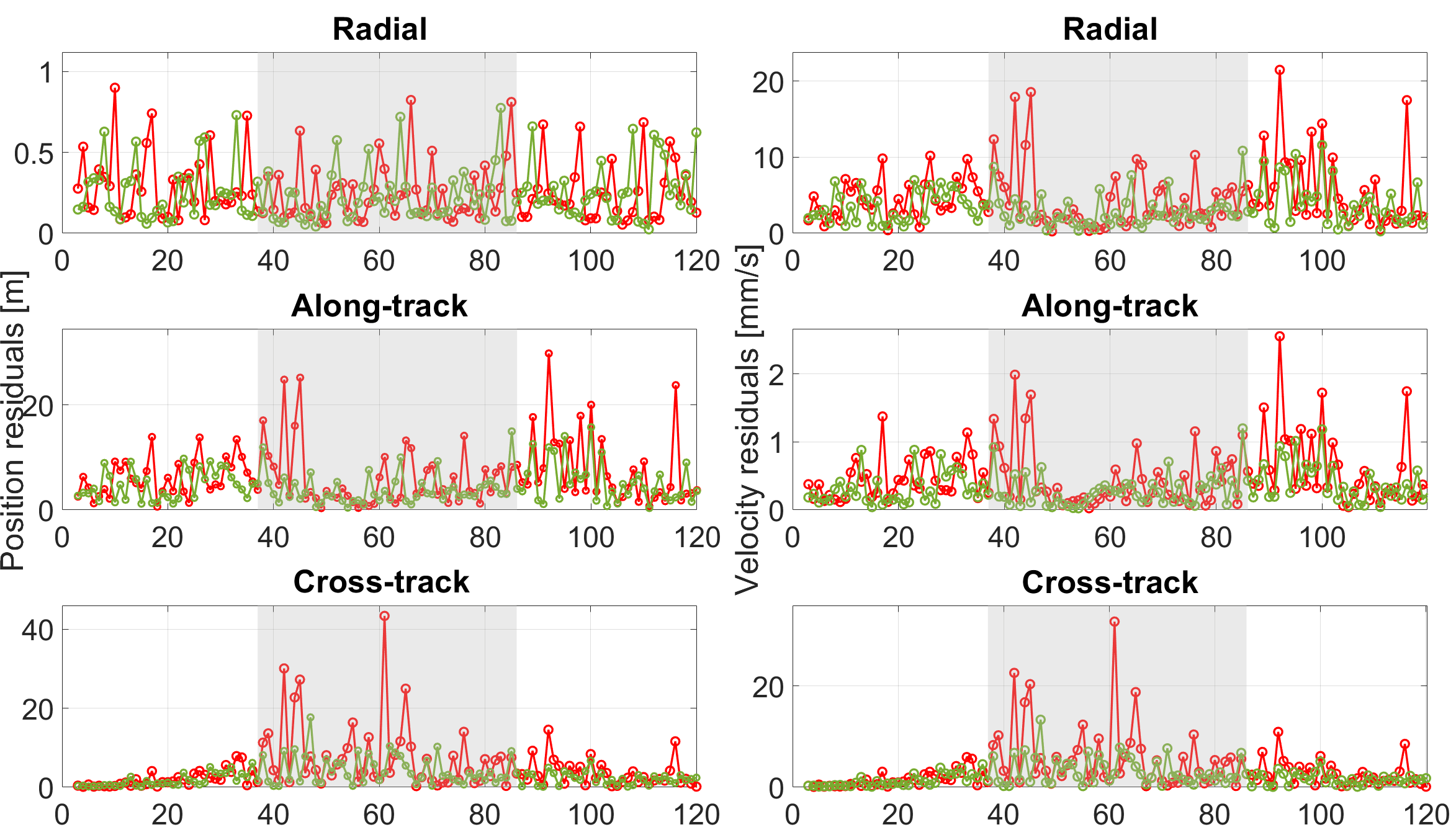}
    \caption[Position and velocity errors vs time]{
    Position and velocity errors for the first $120$ days of the mission. Orbits are propagated from our final converged solution using one year of Doppler observations, when assuming an optimistic (green) or a more realistic (red) scenario for the accelerometer noise.
    Considerably larger errors in the cross-track direction are observed between days $30$ and $70$, corresponding to periods when the $\beta$-Earth angle (as defined in Section~\ref{sec:modelingDop}) consistently remains below $45^\circ$ (gray area).}
    \label{fig:orbRec}
       \end{adjustwidth}
\end{figure}

\subsection{Impact of Observation Geometry on Orbit Determination and Accelerometer Bias Recovery}

Figure~\ref{fig:bias_beta} illustrates the estimation error of the accelerometer bias as a function of the $\beta$-Earth angle. To emphasize the dependency, a bar plot of binned $3\sigma$ error is superimposed on the scatter values. It can be observed that the cross-track bias is strongly correlated with the $\beta$-Earth angle: large errors occur whenever $\beta$-Earth falls below $35^\circ$, which is consistent with the poor Doppler observation geometry in this regime~\citep[see, \textit{e.g.},][]{desprats2024thesis}. These errors in cross-track bias recovery directly propagate into orbit determination errors. In contrast, the along-track bias exhibits only a slight linear increase with $\beta$-Earth angle, but the effect is negligible, since the error remains about two orders of magnitude smaller than in the cross-track error. The radial bias shows no noticeable dependency on $\beta$-Earth.

\begin{figure}[htbp]
\begin{adjustwidth}{-\extralength}{0cm}
\centering
    \includegraphics[width=19.0cm]{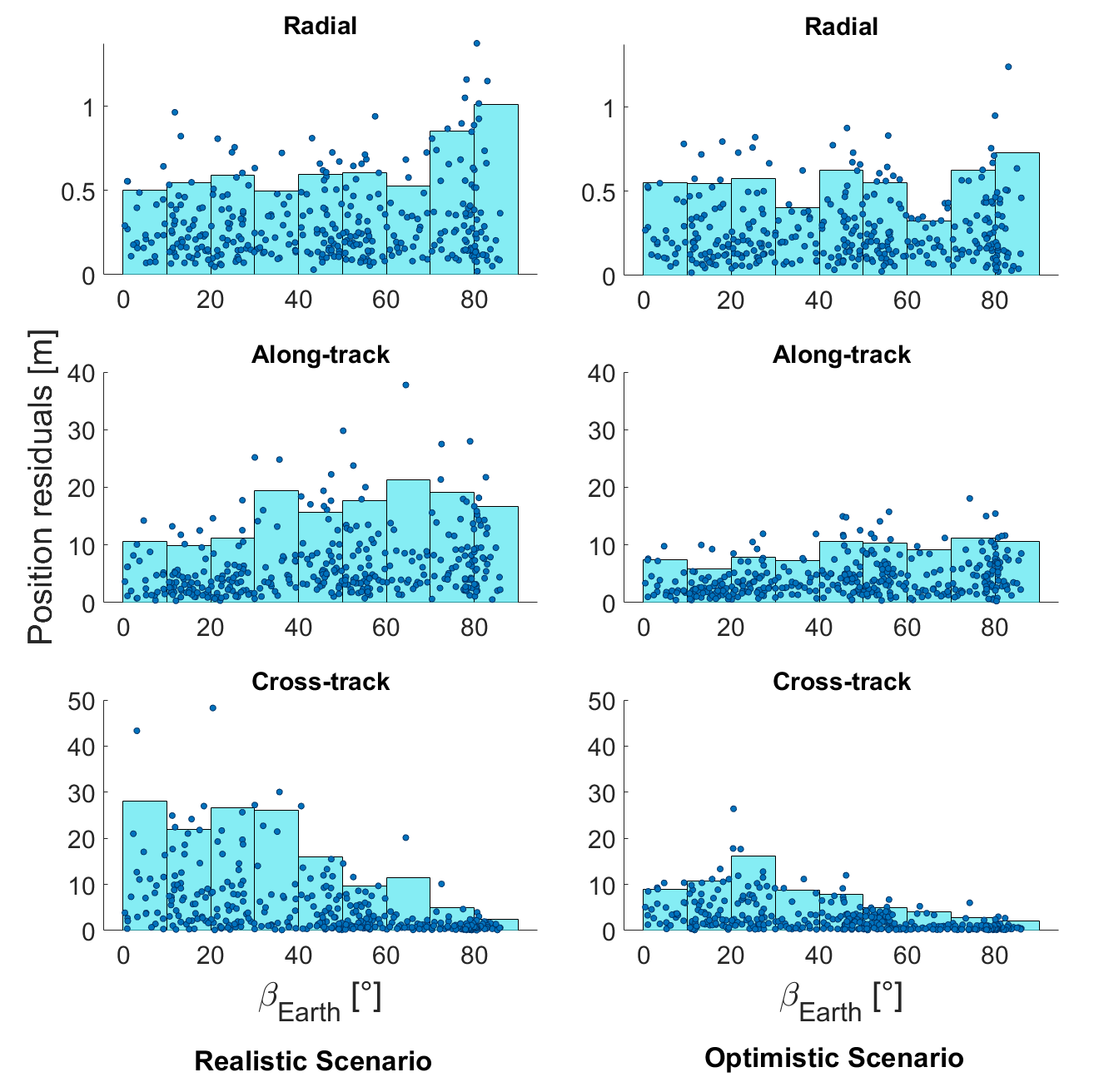}
    \caption[Orbit residuals RMS vs beta angle]{Orbit residuals in the radial (top), along-track (middle), and cross-track (bottom) directions as a function of the $\beta$-Earth angle. $3\sigma$ bins are overlaid on the scatter values to highlight the dependency on the $\beta$-Earth angle. The left column shows the realistic scenario, while the right column shows the optimistic scenario.
}
    \label{fig:orbResBeta}
       \end{adjustwidth}
\end{figure}

\begin{table}[H]
\caption[Position determination accuracy before and after bias est]{Accuracy of spacecraft position determination at different directions, before and after co-estimating the accelerometer biases. The results are based on one year of Doppler observations. \label{posResRSWTab}}
	\begin{adjustwidth}{-\extralength}{0cm}
		\begin{tabularx}{\fulllength}{lCCC}
			\toprule
			\multirow{2}{*}{\textbf{Scenario}}	& \textbf{ RMS error in}	& \textbf{RMS error in}     & \textbf{RMS error in}\\
            				& \textbf{ radial direction   }	& \textbf{along-track direction  }     & \textbf{cross-track direction  }\\
			\midrule
\multirow{2}{*}{Before co-estimation of accelerometer bias }            & \multirow{2}{*}{$0.2~m$}  & \multirow{2}{*}{$98.7~m$}   & \multirow{2}{*}{$131.3~m$}   \\
  &   &  &   \\
\midrule
\multirow{2}{*}{After bias estimation - Realistic scenario}            & \multirow{2}{*}{$0.3~m$}  & \multirow{2}{*}{$8.4~m$}   & \multirow{2}{*}{$8.1~m$}   \\
  &   &  &   \\
\midrule
\multirow{2}{*}{After bias estimation  - Optimistic scenario}            & \multirow{2}{*}{$0.3~m$}  & \multirow{2}{*}{$5.4~m$}   & \multirow{2}{*}{$4.3m$}   \\
  &   &  &   \\
 			\bottomrule
 		\end{tabularx}
 	\end{adjustwidth}
 \end{table}

 \begin{table}[H]
 \caption[Velocity determination accuracy before and after bias est]{Accuracy of spacecraft velocity determination at different directions, before and after co-estimating the accelerometer biases. The results are based on one year of Doppler observations.\label{velResRSWTab}}
 	\begin{adjustwidth}{-\extralength}{0cm}
 		\begin{tabularx}{\fulllength}{lCCC}

 			\toprule
						\multirow{2}{*}{\textbf{Scenario}}	& \textbf{ RMS error in}	& \textbf{RMS error in}     & \textbf{RMS error in}\\
            				& \textbf{ radial direction   }	& \textbf{along-track direction  }     & \textbf{cross-track direction  }\\
			\midrule
            \multirow{2}{*}{Before co-estimation of accelerometer bias}             & \multirow{2}{*}{$71.4~mm/s$}  & \multirow{2}{*}{$8.7~mm/s$}   & \multirow{2}{*}{$99.2~mm/s$}   \\
  &   &  &   \\
\midrule
\multirow{2}{*}{After bias estimation - Realistic scenario}             & \multirow{2}{*}{$6.1~mm/s$}  & \multirow{2}{*}{$0.7~mm/s$}   & \multirow{2}{*}{$6.1~mm/s$}   \\
  &   &  &   \\
\midrule
\multirow{2}{*}{After bias estimation - Optimistic scenario}            & \multirow{2}{*}{$3.9~mm/s$}  & \multirow{2}{*}{$0.4~mm/s$}   & \multirow{2}{*}{$3.2~mm/s$}   \\
 &   &  &   \\
			\bottomrule
		\end{tabularx}
	\end{adjustwidth}
\end{table}

\begin{table}[H]
\caption[Orbit-determination accuracy vs beta angle]{Accuracy of spacecraft orbit determination at different $\beta$-Earth angles. All solutions are obtained after co-estimating the accelerometer bias.\label{orbResTab}}
	\begin{adjustwidth}{-\extralength}{0cm}
		\begin{tabularx}{\fulllength}{lCCC}
			\toprule
			\multirow{2}{*}{\textbf{Scenario}}	& \textbf{ RMS error when}	& \textbf{RMS error when}     & \textbf{RMS error}\\
            				& \textbf{ $\boldsymbol{\beta-Earth~<45~\deg}$ }	& \textbf{$\boldsymbol{\beta-Earth~>45~\deg}$}     & \textbf{total}\\
			\midrule
\multirow{2}{*}{Position - Realistic scenario}            & \multirow{2}{*}{$13.8~m$}  & \multirow{2}{*}{$10.1~m$}   & \multirow{2}{*}{$12.4~m$}   \\
  &   &  &   \\
\midrule
\multirow{2}{*}{Position - Optimistic scenario}            & \multirow{2}{*}{$7.4~m$}  & \multirow{2}{*}{$6.5~m$}   & \multirow{2}{*}{$6.9~m$}   \\
  &   &  &   \\


			\midrule
\multirow{2}{*}{Velocity - Realistic scenario}             & \multirow{2}{*}{$10.2~mm/s$}  & \multirow{2}{*}{$7.3~mm/s$}   & \multirow{2}{*}{$8.6~mm/s$}   \\
  &   &  &   \\
\midrule
\multirow{2}{*}{Velocity - Optimistic scenario}            & \multirow{2}{*}{$5.5~mm/s$}  & \multirow{2}{*}{$4.8~mm/s$}   & \multirow{2}{*}{$5.1~mm/s$}   \\
 &   &  &   \\
			\bottomrule
		\end{tabularx}
	\end{adjustwidth}
\end{table}


We use the final recovered gravity field, obtained from one year of Doppler observations in the nominal mission, to propagate the post-fit orbit.
Figure~\ref{fig:orbRec} displays the position and velocity residuals of our solution for the MPO orbit over a 120-day period, under different assumptions regarding accelerometer noise. 
The centimetre-level accuracy achieved in the radial component and the metre-level accuracy in the along-track and cross-track directions
align with previous studies, as discussed in Section~\ref{sec:intro}. 

Tables~\ref{posResRSWTab} and~\ref{velResRSWTab} show the orbit residuals in different directions, before and after co-estimating the accelerometer biases. When the accelerometer bias is co-estimated together with the orbit and the gravity field coefficients, the resulting errors are considerably reduced, leading to more precise orbit determination.

Figure~\ref{fig:orbResBeta} presents the orbit determination error in different directions as a function of the $\beta$-Earth angle, with bar plots of binned $3\sigma$ error overlaid on the scatter plots to highlight the distribution of variations. A clear dependence is observed in the cross-track component, where the error increases significantly for $\beta$-Earth angles below $45^\circ$. This behavior is similar to that found for the recovery of the cross-track bias (see Figure~\ref{fig:bias_beta}) and reflects the unfavorable Doppler observation geometry at low $\beta$-Earth values. The along-track component shows only a weak trend of increasing error with $\beta$-Earth angle, while the radial component exhibits no obvious sensitivity.
Similarly, in Figure~\ref {fig:orbRec}, significantly larger errors in the cross-track direction are evident between days $30$ to $70$, coinciding with periods when the $\beta$-Earth angle persistently stays below $45^\circ$. 

Table~\ref{orbResTab} 
compares the total accuracy of spacecraft position and velocity determination (combined over all axes) under different assumptions for the accelerometer random noise and under different orbit configurations, and demonstrate the dominant impact of cross-track errors on the overall orbit determination accuracy. Considering the observation geometry, the full orbit becomes visible from Earth once the $\beta$-Earth angle exceeds roughly $38^\circ$–$56^\circ$ (depending on spacecraft altitude); above this threshold the spacecraft is not occulted by the planet. This range corresponds well to the $\sim 45^\circ$ threshold observed in the cross-track (see Figure~\ref{fig:orbResBeta}) and overall orbit determination errors.

\section{Discussion} \label{sec:diss}

The Mercury Planetary Orbiter (MPO), one of the two spacecraft in the ESA/JAXA BepiColombo mission, will play a key role in studying Mercury's gravity field, internal dynamics, and surface properties. 
This study investigates the influence of estimating accelerometer parameters relevant to the onboard Italian Spring Accelerometer (ISA) on orbit and gravity field recovery under varying observation geometries. It also highlights key challenges involved in the simultaneous estimation of orbit, gravity, and accelerometer parameters, and evaluates promising strategies to address these challenges.
We first propagated the MPO orbit using a realistic force model to generate synthetic 2-way Ka-band Doppler observations. A degraded force model (including a truncated gravity field and noisy accelerometer measurements) and errors in the initial state vector were assumed as the a priori knowledge. Orbits were propagated and estimated in daily arcs, iteratively solving for parameters including the initial state vector, gravity field coefficients, and accelerometer parameters. 


Our gravity recovery simulations highlight the importance of carefully selecting arcs during the initial iterations, as a significant number of arcs with large Doppler and position residuals can degrade the solution. By eliminating over $50\%$ of arcs after the first iteration, we were able to enhance the gravity field recovery, even when starting with a poorly known (only up to d/o 8) a priori field. We found that excluding accelerometer parameters in the early iterations mitigates bias–gravity correlations and prevents them from degrading the gravity field solution. 
Once a preliminary gravity field solution was established, accelerometer biases were co-estimated to further improve the solution.
Additionally, we concluded that the
daily estimation of accelerometer biases is crucial for absorbing the remaining mismodeling in the gravity field and achieving an accurate orbit and gravity field solution. We also learned that improving the gravity field without estimating accelerometer biases in the early iterations is critical for the successful recovery of accelerometer biases in later stages.

Our findings demonstrate the potential for achieving gravity field recovery up to degree and order $40$, using one year of 2-way Doppler observations and applying Kaula regularization. 
We observed improvements in the recovery of lower-degree components (d/o $2$ to $12$) when assuming more optimistic accelerometer noise levels. 
Additionally, we found that the estimation error of cross-track bias is highly correlated with the $\beta$-Earth, with errors increasing by an order of magnitude when the $\beta$-angle falls below $45^\circ$, where Doppler observations become less effective in constraining the orbit. During these periods, large errors in cross-track bias recovery are observed, as expected. 
Our results demonstrate that the MPO orbit can be recovered with centimeter-level accuracy in the radial direction and meter-level accuracy in the along-track and cross-track directions, consistent with previous studies. A noticeable increase in cross-track errors is observed at low $\beta$-Earth angles, while the along-track component shows only a slight increase and the radial component remains unaffected. This emphasizes the strong influence of observation geometry in shaping the error behavior.
 
The focus of the paper lies in evaluating promising strategies for the combined recovery of the gravity field, spacecraft orbit, and accelerometer characteristics, rather than emphasizing final accuracy metrics. 
The findings lay the groundwork for more advanced investigations under a broader range of assumptions and more realistic mission scenarios. Future work will also examine additional factors such as the sensitivity of the solution to arc length. Overall, this study demonstrates the critical role that accelerometer parametrization plays in shaping the quality of both orbit determination and gravity field recovery.

\vspace{6pt} 

\funding{This study has been supported by the Swiss National Science Foundation (SNSF) grant $\#200021\_185056$ ``Callisto geodesy: A simulation study to support further space missions to the Jovian system". SB acknowledges support by NASA under award number 80GSFC24M0006.}

\acknowledgments{We gratefully acknowledge the valuable support of our colleague Roberto Peron (Istituto di Astrofisica e Planetologia Spaziali, IAPS-INAF) for his input on the modeling of the accelerometer noise. Calculations were performed on UBELIX, the HPC cluster at the University of Bern (http://www.id.unibe.ch/hpc).}

\conflictsofinterest{The authors declare no conflicts of interest. The funders had no role in the design of the study; in the collection, analyses, or interpretation of data; in the writing of the manuscript; or in the decision to publish the results.
} 



\abbreviations{Abbreviations}{
The following abbreviations are used in this manuscript:
\\

\noindent 
\begin{tabular}{@{}ll}
MPO & Mercury Planetary Orbiter\\
ISA &  Italian Spring Accelerometer\\
MORE & Mercury Orbiter Radio-Science Experiment\\
BELA & BepiColombo Laser Altimeter \\
IR & Infrared Radiation\\
d/o & degree and order \\
ESA & European Space Agency\\
JAXA & Japan Aerospace Exploration Agency\\
\end{tabular}
}




\begin{adjustwidth}{-\extralength}{0cm}

\reftitle{References}


\bibliographystyle{unsrtnat}  

\bibliography{main}

\end{adjustwidth}
\end{document}